\documentclass[twocolumn,showpacs,pre,aps]{revtex4}

\usepackage{dcolumn}
\usepackage{bm}
\usepackage{footnote}

\begin{document}

\title{On statistics and 1/f noise of molecular
random walk in low-density gas}

\author{Yuriy E. Kuzovlev}
\email{kuzovlev@fti.dn.ua} \affiliation{Donetsk Institute for Physics
and Technology, 83114 Donetsk, Ukraine}

\date{13 July 2010}

\begin{abstract}
The random walk of test particle in low-density gas is considered
basing on approximate coarsened version of the collisional
representation of the BBGKY equations. The coarsening presumes that
momentum relaxation rates of the test particle and gas atoms are
equal but allows to analyze the case when their masses are different.
It is shown that both the spectrum exponent and probability
distribution of 1/f-type diffusivity fluctuations of the test
particle essentially depend on ratio of the masses, and corresponding
distribution of its path is found.
\end{abstract}


\maketitle

\section{Introduction}

In \cite{i1,i2} and \cite{p1} it was argued (see also
\cite{tmf,artmf,p3,jst,hs,hs1} we argued that kinetics of spatially
non-uniform gas does not undergo the Boltzmann equation (or other
related and similar equations) even under the low-density or
Boltzmann-Grad limit ($\,\nu\rightarrow\infty\,$,
$\,\delta\rightarrow 0\,$, $\,\nu\delta^2=\,$const\,, with $\,\nu\,$
being concentration of gas particles and $\,\delta\,$ radius of their
repulsive interaction). Instead, an infinite chain of kinetic
equations appears - for the one-particle distribution function (DF)
plus infinite chain of specific two-particle, three-particle, etc.
DFs which represent mean density of pair collisions, density of two
connected pair collisions, and so on. Solution of such chains (as
well as their rigorous formulation) will be a great challenge for
future theory. However, instructive and useful approximate solutions
dan obtained already at present \cite{i1,i2,p1,jst,p2}, first of all,
for situations when the system (gas or liquid) as a whole is almost
equilibrium, and the spatial non-uniformity is rather of statistical
than thermodynamical nature. For example, when one considers random
walk (``Brownian motion'') of a test (or probe, or marked) particle
in a fluid.

In \cite{i1,i2} and later in \cite{p1,p2} we considered (at different
approximations) the case when the test particle, or ``Brownian
particle'' (BP), is one of gas atoms. Here we consider the case when
BP differs from gas atoms (e.g. is an impurity atom).

\section{Basic equations}

We star from the Bogolyubov-Born-Green-Kirkwood-Yvon (BBGKY)
equations \cite{bog} for a molecule (molecular-size ``Brownian
particle'') dissolved in a fluid \cite{tmf,artmf,p3,jst}:
\begin{eqnarray}
\frac {\partial F_n}{\partial t}=[\,H_{n}\,,F_n\,]\,+\,\nu\, \frac
{\partial }{\partial {\bf P}}\int_{n+1}\!\! \Phi^{\,\prime}_b({\bf
R}-{\bf r}_{n+1})\,F_{n+1}\,+\,\nonumber\\
+\,\nu \sum_{j\,=1}^n\,\frac {\partial }{\partial {\bf
p}_j}\int_{n+1}\!\!\Phi^{\,\prime}_a({\bf r}_j-{\bf r}_{n+1})
\,F_{n+1}\, \,\,,\,\,\, \,\, \,\,\,\, \label{fn}
\end{eqnarray}
where $\,n\,=\,0,\,1,\,\dots\,$,\,\,,\,
$\,\Phi^{\,\prime}_{a,\,b}({\bf r})=\nabla\Phi_{a,\,b}({\bf r})\,$\,
are ``atom-atom'' and ``BP-atom'' interaction forces, and all
designations are the same as in \cite{tmf} or \cite{artmf}. Since we
here again are interested in BP's random walk in (quasi-)equilibrium
fluid, the initial conditions will be
\begin{equation}
\begin{array}{c}
F_n(t=0)\,=\, \Delta({\bf R}-{\bf R}_0)\,F_n^{(eq)}({\bf
r}^{(n)}\,|{\bf R};\nu)\,\times \\
\times\,G_M({\bf P})\prod_{j\,=1}^n G_m({\bf p}_j)\,\,\,,\label{ic}
\end{array}
\end{equation}
where\, $\,F_n^{(eq)}({\bf r}^{(n)}\,|{\bf R};\nu)\,$ are usual
thermodynamically equilibrium DF for $\,n\,$ atoms in presence of BP
occupying point $\,{\bf R}\,$\,,\,
\[
G_m({\bf p})\,=\,(2\pi Tm)^{-\,3/2}\exp{(-{\bf p}^2/2Tm)}\,\,
\]
is the Maxwell momentum distribution of a particle with mass
$\,m\,$\,,\, and $\,\Delta(\rho)\,$ is a normalized probability
distribution like a``smoothed delta-function''. We then want to
reveal how probability distribution of BP's position $\,\bm{R}\,$
evolves after start from $\,\Delta(\bm{R}-\bm{R}_0)\l$.

\section{Collisional representation}

If our fluid is a sufficiently dilute gas, or the Boltzmann-Grad gas,
then inter-particle interactions can be described in terms of well
separated collisions. Since any ``collision'' is a finite duration
process, while the distribution functions (DF) $\,F_n\,$ describe
instant states of the system, we should reformulate DFs and equations
(\ref{fn}) in terms of collision processes.

In principle, this is very simple task. First of all, let us consider
the $\,(n+1)$-particle Poisson bracket, $\,[H_n,F_n]\,$, describing
separate evolution of  a group of $\,n\,$ fluid atoms plus BP, and
divide it into two parts, one of which describes drift of the center
of mass of the group while another relative displacement of particles
inside the group:
\begin{equation}
[H_n,F_n]\,=-\bm{V}_n\cdot \frac {\partial F_n}{\partial
\bm{Z}_n}-\frac {\partial F_n}{\partial \Theta_n}\, \label{div}
\end{equation}
We introduced the center of mass positions and velocities,
\[
\begin{array}{c}
{\bf Z}_n=\frac {M{\bf R}+m\sum_{j=1}^n {\bf r}_j}{M+nm}\,\,,
\,\,\,\,\, {\bf V}_n= \frac {{\bf P}+\sum_{j=1}^n {\bf
p}_j}{M+nm}\,\,,
\end{array}
\]
and (for $\,n>0\,$) the inner time of the relative motion,
$\,\Theta_n\,$. For example, at $\,h=1\,$,
\begin{eqnarray}
-\frac {\partial F_1}{\partial \Theta_1}\,=\,-\,\bm{u}_{1}\cdot \frac
{\partial F_1}{\partial \rho_1} +\,
\Phi^{\,\prime}_b(\rho_1)\cdot\left[ \frac {\partial F_1}{\partial
{\bf p}_1} -\frac {\partial F_1}{\partial {\bf P}}\right] \,\,\,
\nonumber 
\end{eqnarray}
with\, $\,\rho_1=\bm{r}_1-\bm{R}\,$\, and\,
$\,\bm{u}_{1}=\bm{v}_1-\bm{V}\,$.

Second, let a given  relative disposition of the particles can be
qualified as a section of collision process (pair collision at
$\,n=1\,$ or sequence of $\,n\,$ connected pair collisions).
 Then it is reasonable to treat corresponding DF
$\,F_n\,$ as a measure of mean (ensemble average) density of such
processes, and consider it as a function of $\,t,{\bf Z}_n, {\bf
V}_n,\Theta_n\,$,\, $\,n\,$ relative incoming ({\it \,in\,}-state)
velocities (e.g. $\,\bm{u}_{j}=\bm{v}_j-\bm{V}\,$) and $\,3n-1\,$\,
input geometric parameters of the process.

Third, however, such the treatment can not be literal: if
$\,F_n(t,\Theta_n,...)\,$ at time moment $\,t\,$ represents ``density
of collisions'' at this time, then its value at $\,t+\Delta t\,$ is
not $\,F_n(t+\Delta t,\Theta_n,...)\,$ but $\,F_n(t+\Delta
t,\Theta_n+\Delta t,...)\,$\,, since the same collision is
represented by different $\,\Theta_n\,$ values at different times.

This statement means that if DF $\,F_n\,$ really describe collisions
as the whole processes then each of Eqs.\ref{fn} for them (at
$\,n>0\,$ must divide into two equations:
\begin{eqnarray}
\frac {\partial F_n}{\partial t}= -\bm{V}_n\cdot \frac {\partial
F_n}{\partial \bm{Z}_n}\,+\,\nu\, \frac {\partial }{\partial {\bf
P}}\int_{n+1}\!\! \Phi^{\,\prime}_b({\bf
R}-{\bf r}_{n+1})\,F_{n+1}\,+\,\nonumber\\
+\,\nu \sum_{j\,=1}^n\,\frac {\partial }{\partial {\bf
p}_j}\int_{n+1}\!\!\Phi^{\,\prime}_a({\bf r}_j-{\bf r}_{n+1})
\,F_{n+1}\, \,\,,\,\,\, \,\, \,\,\,\, \label{cfn}\\
\frac {\partial F_n}{\partial \Theta_n}\,=\,0\,\,\,, \,\,\, \,\,
\,\,\,\,\, \,\, \,\, \,\,\, \,\, \,\,\,\,\, \,\, \,\, \label{tn}
\end{eqnarray}
so that Eq.\ref{tn} must be satisfied ``inside collisions'' (i.e.
space-time regions assigned, in relative coordinates, to a collision
process) while Eq.\ref{cfn} ``out of collisions''. It is natural:
since relative movement of colliding particles is included, along
with their interaction, into collisions, it is excluded from
description of collision events as the wholes (for details and more
explanations, see \cite{i1,i2,p1,jst}).

Fourth, the Eqs.\ref{tn} are the tool for transforming the
interaction integrals in Eq.\ref{cfn} into usual Boltzmannian
``collision integrals'':
\begin{eqnarray}
\frac {\partial F_n}{\partial t}\,=\,-\bm{V}_{n}\cdot\frac {\partial
F_n}{\partial \bm{Z}}\,+\, \nu\, \widehat{S}_{n+1}^{\,b}\,
F_{n+1}^{\,in}\,+ \nonumber\\
+\, \nu\sum_{j=1}^n \widehat{S}_{j\,n+1}^{\,a}F_{n+1}^{\,in}\,\,\,,\,
\label{bi}
\end{eqnarray}
where $\,\widehat{S}_{j\,k}^{\,a}\,$ and
$\,\widehat{S}_{k}^{\,b}\,$\, are the Boltzmann collision operators
acting on $\,j$-th and $\,k$-th atoms' velocities and on BP's and
$\,k$-th atom velocities, respectively, and DFs
$\,F_{n+1}^{\,in}\,$\, describe such $\,(n+1)$-particle
 configurations when the ``outer'' $\,(n+1)$-th particle is
just on the border of the collision.

Such the border configurations and corresponding DFs need in special
consideration \cite{i1,i2,p1}. If the most (infinite) part of the gas
stay in thermodynamic equilibrium during all (finite-time) evolution
of initial (finite-range) disturbance of the equilibrium, then
naturally
\begin{eqnarray}
F_{n+1}^{in}\,=\,G_m(\bm{p}_{n+1})\, \int F_{n+1}\,
d\bm{p}_{n+1}\,\,, \label{mch}
\end{eqnarray}
and Eqs.\ref{bi} transform into
\begin{eqnarray}
\frac {\partial F_n}{\partial t}\,=\,-\bm{V}_{n}\cdot\frac {\partial
F_n}{\partial \bm{Z}}+ \,\nu \widehat{\Lambda}^b \int
F_{n+1}\,d\bm{p}_{n+1}\,
+\,\nonumber \\
+\, \nu\sum_{j=1}^n \widehat{\Lambda}_j^a\int F_{n+1}\bm{p}_{n+1}\,
\label{cfn}
\end{eqnarray}
with $\,\widehat{\Lambda}_j^{a,b}\,$\, being the Boltzmann-Lorentz
operators. The latter are defined by
\begin{eqnarray*}
\widehat{S}_{j\,n+1}^{\,\,a}\,[F(...\bm{p}_j...)\,G_m(\bm{p}_{n+1})]\,=
\,\widehat{\Lambda}_j^a\,
F(...\bm{p}_j...)\\
\widehat{S}_{n+1}^{\,\,b}\,[F(...\bm{P}...)\,G_m(\bm{p}_{n+1})]\,=
\,\widehat{\Lambda}^b\, F(...\bm{P}...)
\end{eqnarray*}
Relations (\ref{mch}) and hence Eqs.\ref{cfn} do express the
``molecular chaos'', but with those principal difference from the
Boltzmann's one that here it concerns velocities of colliding
particles only. What is for coordinates of the particles, they
possess mutual statistical correlations, as far strong as strong is
non-uniformity of the coordinates probability distributions (more
comments see \cite{i1,i2,p2}).

Of course, the transition from Eqs.\ref{fn} to Eqs.\ref{cfn} should
be accompanied by corresponding transition from initial conditions
(\ref{ic}) to their collisional representation:
\begin{eqnarray}
F_n(t=0,\bm{Z},\bm{P},\bm{p}^{(n)}) \,=\,\,\,\,\,\,
\,\, \,\,\,
\,\, \, \,\,\,\,\, \, \,\,\,\,\, \,\nonumber\\
=\int_{\Omega^h} \int\,\,\delta\left(\bm{Z}-\frac
{M\bm{R}+m\sum_{j=1}^n\bm{r}_j}{M+nm}\right)\,\times\nonumber\\
\times\,F_n(t=0,\bm{R},\bm{r}^{(n)},\bm{P},\bm{p}^{(n)})\,
d\bm{R}\,\prod_{j=1}^n \frac
{d\bm{r}_j}{\Omega}\,= \nonumber\\
=\, \left(1+\frac {nm}{M}\right) \int_{\Omega^n} \Delta\left(\bm{Z}-
\frac mM \sum_{j=1}^n \rho_j-\bm{R}_0\right)\, \times\nonumber\\
\times\, \,\prod_{j=1}^n \frac {d\rho_j}{\Omega}\,\,
\prod_{j=1}^nG_m(\bm{p}_j)\,\,\,, \nonumber
\end{eqnarray}
where $\,\rho_j=\bm{r}_j-\bm{Z}\,$ and $\,\Omega\,$ is a ``collision
volume'', i.e. volume of a region assigned (in each of the spaces
$\,\rho_j=\bm{r}_j-\bm{Z}\,$) per one collision. Thus under the
Boltzmann-Grad limit, when a width of the distribution
$\,\Delta(\rho)\,$ is krpt constant in units of $\,\lambda\,$ we can
write, obviously,
\begin{eqnarray}
F_n(t=0,\bm{Z},\bm{P},\bm{p}^{(n)}) \,=\,\,\,\,\,\, \,\, \,\,\, \,\,
\, \,\,\,\,\, \, \,\,\,\,\, \,\nonumber\\
=\,\left(1+\frac {nm}{M}\right)\,\Delta(\bm{Z}-\bm{R}_0)\,
\prod_{j=1}^nG_m(\bm{p}_j)\, \label{cic}
\end{eqnarray}
This is generalization of initial conditions derived in \cite{p1},
directly from the Gibbs canonical ensemble, for the case when BP is
merely one of gas atoms.

Notice also that under the Boltzmann-Grad limit all the variety of
the center of mass coordinates $\,\bm{Z}_n\,$ can be replaced by
single common variable  $\,\bm{Z}\,$.

\section{Diffusive evolution and more simplification of equations}

The Eqs.\ref{cfn} still seem very complicated. Therefore we would
like to further coarsen and simplify them being eventually interested
in the distributions in configurational spaces,
\begin{eqnarray*}
W_n(t,\bm{Z})\,=\int\!...\!\int F_n
\,d\bm{p}_1...d\bm{p}_n\,d\bm{P}\, \,
\end{eqnarray*}
Since, according to Eqs.\ref{cfn},
\begin{eqnarray}
\frac {\partial W_n}{\partial t}\,=\,-\frac {\partial
\bm{Q}_n}{\partial \bm{Z}}\,\,\,, \label{cwn}
\end{eqnarray}
where $\,\bm{Q}_n\,$ are corresponding probability flows,
\begin{eqnarray}
\bm{Q}_n(t,\bm{Z})\,=\int\!...\!\int \bm{V}_n\,
F_n\,d\bm{p}_1...d\bm{p}_n\,d\bm{P}\,\,\,,\nonumber
\end{eqnarray}
we have to consider also at least evolution equation for these flows,
\begin{eqnarray}
\frac {\partial \bm{Q}_n}{\partial t}\,=\,-\frac {\partial }{\partial
\bm{Z}}\int\! ...\! \int {\bf V}_n\,{\bf V}_n\,F_n\, \,d{\bf p}_1
...d{\bf p}_n\,d{\bf P} +\,\,\,\,\, \label{qe}\\
+ \int\! ...\! \int {\bf V}_n\,\left[
\widehat{\Lambda}^b+\sum_{j=1}^n
\widehat{\Lambda}^a_j\,\right]\,F_{n+1}\, \,d{\bf p}_1 ...d{\bf
p}_{n+1}\,d{\bf P}\,\nonumber
\end{eqnarray}

In order to express right-hand sides here via $\,\bm{Q}\,$'s and
$\,W_n\,$'s only and thus close the equations, we need in two
assumptions. The first is that the conditional velocity
distributions, $\,F_n/W_n\,$, duffer from the equilibrium ones only
by non-zero ($\,\bm{Z}\,$-dependent) mean drift velocities,
$\,\overline{\bm{V}}_n=\bm{Q}_n/W_n],$l which is common for all the
velocities ($\,\bm{V}\,$ and $\,\bm{v}_j\,$) under consideration.
This hypothesis, in turn, can be correct only under the second
assumption, namely, that all the velocities possess equal relaxation
rates. Formally, the equality
\[
\bm{V}_n\,\left[\widehat{\Lambda}^b+\sum_{j=1}^n
\widehat{\Lambda}^a_j\,\right]\,=\,-\,\gamma\,\bm{V}_n\,\,\,
\]
approximately takes place, that is $\,\bm{P}=M\bm{V}\,$ and
$\,\bm{p}_j=m\bm{v}_j\,$ are left eigenfunctions of operators
$\,\widehat{\Lambda}^b\,$ and $\,\widehat{\Lambda}^a_j\,$,
respectively, corresponding to the same eigenvalue $\,-\gamma\,$.
Then, taking into account also  the initial conditions (\ref{cic}),
we can expect that
\begin{eqnarray} \int\! ...\! \int ({\bf
V}_n)_i\,({\bf V}_n)_j\,\,F_n\, \,d{\bf p}_1 ...d{\bf p}_n\,d{\bf
P}\,\approx\, \frac {T\,\delta_{ij}}{M+nm}\,W_n\, \,,\,\, \nonumber
\end{eqnarray}
as in equilibrium. Consequently Eqs.\ref{qe} reduce to
\begin{eqnarray}
\frac {\partial\bm{Q}_n}{\partial t}\,=\,-\frac {T}{M+nm}\,\frac
{\partial W_n}{\partial \bm{Z}}
-\gamma\,\bm{Q}_{n+1}\,\,\,\,\,\label{cqn}
\end{eqnarray}
and close Eqs.\ref{cwn}.

Of course, on one hand, the second of our assumptions looks rather
unnatural, since in reality operators $\,\widehat{\Lambda}^b\,$ and
$\,\widehat{\Lambda}^a_j\,$, may have very different eigenvalues.
But, on the other hand, the corresponding approximation still allows
to consider the case of different masses of BP and atoms, $\,M\neq
m\,$l and thus obtain a generalization of results of \cite{p1}.

Clearly, our approximation reflects the fact that the system ``gas +
BP'' is non-equilibrium in statistical (or informational) sense only
but not in literally thermodynamic sense, and therefore its evolution
 is sooner ``diffusive'' than ``hydrodynamical''. Nevertheless, if
 velocities of BP and atoms relax (and fluctuate) with different rates
then their conditional distribution inside the collisional clusters
can be significantly non-equilibrium (non-Gaussian), and we need in a
more complicated approximation.

For the present let us confine ourselves by the Eqs,\ref{cwn} and
\ref{cqn}. Initial conditions to them, as follow from (\ref{cic}),
are
\begin{eqnarray}
W_n(t=0,\bm{Z})\,=\,\left(1+\frac
{nm}{M}\right)\,\Delta(\bm{Z}-\bm{R}_0)\,\,\,,\label{sic}\\
\bm{Q}_n(t=0,\bm{Z})\,=\,0\,\,\,, \,\,\,\,\,\, \,\,\,\,\,\,
\,\,\,\,\,\,\nonumber
\end{eqnarray}
where now, of course, we can merely $\,\delta(\bm{Z})\,$ in place of
$\,\Delta(\bm{Z}-\bm{R}_0)\,$.

\section{Analysis of shortened equations}

As usually, let us make the Laplace and Fourier transforms, writing
\[
F(p,\bm{k})\,=\, \int_0^\infty\! dt\,\, e^{-p\,t}  \int\! d\bm{Z}\,\,
e^{i\bm{k}\cdot\bm{Z}}\, F(t,\bm{Z})\,\,\,,
\]
and introduce designations
\begin{eqnarray}
 V_0=\sqrt{\frac TM}\,\,, \,\,\,
v_0=\sqrt{\frac Tm}\,\,,\,\,\, D_b=\frac {V_0^2}{\gamma }\,\,,
\,\,\, D_a=\frac {v_0^2}{\gamma }\,\,,\nonumber\\
\alpha = \frac mM\,\,, \,\,\,\, X=\frac {V_0^2\,\bm{k}^2}{p^2}\,\,
\,\,\,,\, \, \,\,\,\, \,\,\,\,\,\,\,\,\,\,\,\,\,\, \,\, \,\,\,\,\,
\,\,\,\, \,\,\,\,\,\, \nonumber
\end{eqnarray}
Then Eqs.\ref{cwn} and \ref{cqn} together with Eq.\ref{sic} yield
\begin{eqnarray}
W_n(p,\bm{k}) = \frac {1+\alpha n}{p}\,+\frac {i\bm{k}}{p}\cdot
\bm{Q}_n(p,\bm{k})\,\,\,, \,\,\,\,\,\,\,\,\,\,\, \label{w}\\
\left\{1+\frac {X}{1+\alpha n} \,\right\} \bm{Q}_n(p,\bm{k}) = \frac
{i\bm{k}\,V_0^2}{p^2} - \frac {\gamma}p \,\bm{Q}_{n+1}(p,\bm{k})
\,\,\, \,\,\label{mij}
\end{eqnarray}
According to characteristic structure of the chain of equations
(\ref{mij}), its solution can be written as sum of infinite iteration
series. This yields
\begin{eqnarray}
\bm{Q}_n = -\frac {i\bm{k}\,V_0^2}{p\gamma} \, \sum_{s=n}^\infty
\prod_{r=n}^s \left(-\frac {\gamma}{p}\cdot \frac {1+\alpha
r}{X+1+\alpha r}\right)\!
\,\,\,,\,\,\,\nonumber 
\\
W_n=\frac {1+\alpha n}{p}+\frac X\gamma \sum_{s=n}^\infty
\prod_{r=n}^s \left(-\frac {\gamma}{p}\cdot \frac {1+\alpha
r}{X+1+\alpha r}\right) \,\,\,\,\, \, \label{erw}
\end{eqnarray}
This is generalization of of the series from \cite{p1}.

Next, let us introduce quantities
\[
a=\frac 1{\alpha} +n\,\,\,, \,\,\,\,\, ,\,\,\, c= \frac X{\alpha}
+\frac 1{\alpha} +n =\frac {v_0^2\bm{k}^2}{p^2} +\frac 1{\alpha}
+n\,\,\,,\,\,\,
\]
and rewrite expression (\ref{erw}) as follows,
\begin{eqnarray}
W_n=\frac {1+\alpha n}{p}+\frac X\gamma \sum_{s=1}^\infty
\left(-\frac {\gamma}p\right)^s \frac
{\Gamma(a+s)}{\Gamma(a)\,(s-1)!}\,B(c,n)\,= \,\nonumber\\
=\frac {1+\alpha n}{p}+\frac X\gamma \sum_{s=1}^\infty  \frac
{(-\,\gamma/p)^s}{\Gamma(a)\,(s-1)!} \int_0^\infty\!\!
x^{a+s-1}\,e^{-x}\,dx\,\times\nonumber\\
\times\, \int_0^1\!\! t^{n-1}\,(1-t)^{c-1}\,dt\,=\,\,\,\,\nonumber\\
=\,\frac {1+\alpha n}{p} \left[1-\frac X\alpha \int_0^1\!\!
\frac{(1-t)^{c-1}}{(1+\gamma t/p)^{a+1}}\,dt\right]
 \nonumber
\end{eqnarray}
Hence,
\begin{eqnarray}
\frac {pW_n}{1+\alpha n}=%
1-\frac X\alpha \int_0^1\!\! \frac{(1-t)^{\,\frac X\alpha + \frac
1\alpha+n-1}}{(1+\gamma t/p)^{\,\frac 1\alpha
+n+1}}\,dt\,\,\,\label{erw1}
\end{eqnarray}
which is direct generalization of formula (27) from \cite{p1}.

\section{Large-scale asymptotic}

As we already know \cite{i1,i2,p1}, most important statistical
characteristics of the BP's random walk what distinguish it from the
usual Ornstein-Uhlenbeck process or similar random processes) are its
fourth-order path statistical moment or cumulant and long-range
asymptotic of the path distribution. The latter is defined by
\begin{eqnarray}
\overline{W}_n(p,\bm{k}) =\,\lim_{\xi\rightarrow
0}\,\xi^2\,W_n(\xi^2 p\,,\,\xi\,\bm{k})\,\,,\nonumber \,\,\\
\overline{\bm{Q}}_n(p,\bm{k})=\,\lim_{\xi\rightarrow 0}\,\xi\,
\bm{Q}_n(\xi^2 p\,,\,\xi\,\bm{k})\,\, \,\,\,\,\,\, \,\,\,\nonumber
\end{eqnarray}
(for details see \cite{p1}). We just foreknow that our random walk
behaves similar;y to diffusion processes (like the Wiener process)
but with randomly varying diffusivity in place of a constant one.
Taking into account results of \cite{p1}, it is reasonable to
interpret the asymptotic in terms of slow (scaleless) diffusivity
fluctuations. Correspondingly, let us write
\begin{eqnarray}
\overline{W}_n = (1+\alpha n)\int_0^\infty \frac {\overline{U}_n(D)\,
dD}{p+D\bm{k}^2}\,\,\,,\label{u}
\end{eqnarray}
where $\,\overline{U}_n(D)\,$ is effective normalized probability
distribution of diffusivity (at $\,n=0\,$ it represents BP's
diffusivity while at $\,n>0\,$ characterizes spreading of densities
of two- and multi-particle collisional events).

In the mentioned limit the expression (\ref{erw1}) turns to
\begin{eqnarray}
\frac {\overline{W}_n}{1+\alpha n}=%
\left[\frac 1\alpha +n+1\right]
\frac 1p \int_0^\infty \frac {\exp{\left(-\,\frac
{pX}{\gamma\alpha}\,y\right)}}
 {(1+y)^{\,\frac 1\alpha +n+2}}\,dy\,\nonumber
 \,\label{erwl}
\end{eqnarray}
or equivalently
\begin{eqnarray}
\frac {\overline{W}_n}{1+\alpha n}=\,\left[\frac 1\alpha +n+1\right]
\!\!\int_0^\infty\frac
{\exp{(-D_a\bm{k}^2\tau)}\,d\tau}{(1+p\tau)^{1/\alpha\,+n+2}}\, \,
\,\,\label{i2}
\end{eqnarray}
Combining it with identity
\[
1/(1+y)^{b+1} = \int_0^\infty x^b\,e^{-\,(1+y)x}\, dx
/\Gamma(b+1)\,\,\,,
\]
with $\,b=1/\alpha +n+1\,$, it is easy to obtain
\[
\frac {\overline{W}_n}{1+\alpha n}=%
\frac {1}{\Gamma(b)} \int_0^\infty \frac {x^b\exp{(-\,x)}\,
dx}{px+D_a\bm{k}^2}\,\,\,\,\,
\]
Comparison of this formula and (\ref{u}) yields
\begin{eqnarray}
\overline{U}_n(D)= \frac {1}{\Gamma(b)\, D} \left(\frac
{D_a}{D}\right)^{b} \exp{\left(-\,\frac {D_a}{D}\right)}\,\,
\label{us}
\end{eqnarray}
with $\,b=1/\alpha +n+1=\frac Mm +n+1\,$.

Hence, the effective long-range diffusivities  are random quantities
with very bad statistics. In particular, its their most probable
(m.p.) and mean values essentially differ one from another:
\[
(\,\texttt{m.\,p.}\,\,\, D\,)_n=\frac {D_a}{2+M/m +n}\,\,,
\,\,\,\,\left\langle D \right\rangle_n =\frac {D_a}{M/m +n}
\]
Notice that $\,\left\langle D \right\rangle_n =D_b\,$.

Making in (\ref{u}) the inverse Laplace and Fourier transforms and
substituting  (\ref{us}), in the space-time representation we have
\begin{eqnarray}
\overline{W}_n(t,\bm{R})\equiv \lim_{s\rightarrow\infty}\,
s^d\,W_n(s^2t,s\bm{R})\,=
\,\,\,\,\,\,\,\,\,\,\,\,\,\,\,\,\label{ngl}\\
= \int_0^\infty (4\pi Dt)^{-d/2} \exp{\left(-\frac
{\bm{R}^2}{4Dt}\right)}\, \overline{U}_n(D)\, dD\,=\,\nonumber\\
\,\,\,=\,\frac {\Gamma (b+d/2)}{(4\pi D_at)^{d/2}\,\,\Gamma(b)\,\,
(1+\bm{R}^2/4D_at)^{\,b+d/2}}\,\,\,,\nonumber
\end{eqnarray}
where\, \,$\,b=1/\alpha +1+n=M/m+1+n\,$\, ,\, $\,\bm{R}\equiv
\bm{Z}\,$\,, and\, $\,d\,$\, is the space dimension ($\,d=3\,$).

\section{Statistical moments and 1/f noise}

Expansion of the $\, W_n(p,\bm{k}) \,$ into series over
$\,\bm{k}^2\,$ gives Laplace transforms of equilibrium statistical
moments of the distributions $\,W_n(t,\bm{R})\,$:
\begin{eqnarray}
\frac {W_n(p,k)}{1+\alpha n}= \frac 1p\!+\!\sum_{s=1}^\infty \frac
{(-k^2)^s}{(2s)!}\int_0^\infty \!\! e^{-p\,t}\,\langle
R^{2s}(t)\rangle_n \,dt\, \,\, \label{sm}
\end{eqnarray}
Here and be;ow  $\,R(t)\,$ means projection of vector $\,\bm{R}(t)\,$
onto arbitrary fixwd axis and
\[
\langle R^{2s}(t)\rangle_n=\int R^{2s}\, W_n(t,\bm{R})\,
d\bm{R}\,\,\,\, \,\,
\]
Direct comparison of (\ref{sm}) with (\ref{erw1}) shows that
\begin{eqnarray}
\int_0^\infty \!\! e^{-p\,t}\,\langle R^{2s}(t)\rangle_n \,dt\,
=\,\frac {(2s)!}{(s-1)!}\cdot \frac 1p  \left(\frac
{v_0^2}{p^2}\right)^s\,\times \,
\,\,\,\,\,\,\label{moms}\\
\times\, \int_0^1\frac {(1-x)^{1/\alpha+n-1}}{(1+\gamma
x/p\,)^{1/\alpha+n+1}}\, \left[\ln{\frac {1}{1-x}} \right]^{s-1}
dx\,\nonumber
\end{eqnarray}

Considering the long-time asymptotic of these expressions, i.e. the
limit $\,p/\gamma \rightarrow 0\,$, we have, firstly,
\begin{eqnarray}
\int_0^\infty \!\! e^{-p\,t}\,\langle R^{2}(t)\rangle_n \,dt
\rightarrow\frac {2V_0^2}{(1+\alpha n) p^2\gamma} =\frac
{2D_b}{(1+\alpha n) p^2}\,\,\, \label{m2s}
\end{eqnarray}
which corresponds to the usual diffusion-law asymptotic\, $\,\langle
R^{2}(t)\rangle_n\rightarrow 2D_b t/(1+\alpha n)\,$\, at \, $\,\gamma
t\rightarrow\infty\,$.

Secondly, an asymptotic behavior of the fourth-order moment crucially
depends on sign of\, $\,M/m+n-1\,$. Namely, if this quantity\,
$\,<1\,$\,, that is\, $n=0\,$\, and\, $\,M/m<1\,$\,, then
\begin{eqnarray}
\int_0^\infty \!\! e^{-p\,t}\,\langle R^{4}(t)\rangle_0 \,dt
\rightarrow\, \frac {24D_b^2}{p^{\,3}}  \left[\frac
{\gamma}{p}\right]^{1-M/m} C\left(\frac Mm\right)\,\,,\,\,\,\,
\label{m4s}\\
C(z)\,\equiv\,z^2\! \int_0^1\frac {(1-x)^{z-1}}{x^{\,z+1}}\,\ln{\frac
{1}{1-x}}\,\,dx\,=\frac {\pi z}{\sin {\pi z}}\,\, \,\,\,\, \,
\,\,\,\, \,\label{cx}
\end{eqnarray}
At\, $n=0\,$\, and\, $\,M/m=1\,$\,, we come to result of \cite{p1},
\begin{eqnarray}
\int_0^\infty \!\! e^{-p\,t}\,\langle R^{4}(t)\rangle_0 \,dt
\rightarrow\, \frac {24D_b^2}{p^{\,3}}   \ln{\frac
{\gamma}{p}}\,\,\label{m40}
\end{eqnarray}
And at\, $\,M/m>1\,$\, under\, $n=0\,$\, we find
\begin{eqnarray}
\int_0^\infty \!\! e^{-p\,t}\,\langle R^{4}(t)\rangle_0 \,dt
\rightarrow\, \frac {24D_b^2}{p^{\,3}}\cdot \frac
{M}{M-m}\,\,\label{m4m}
\end{eqnarray}
These three asymptotics should be compared with \,$\,24D_b^2/p^3\,$\,
which is asymptotic of $\,\langle R^{4}(t)\rangle_0\,$ for the ideal
Gaussian random walk.

What is for the higher-order moments at $\,n=0\,$ and any moments at
$\,n>0\,$, they are presented by table
\begin{eqnarray}
\int_0^\infty \!\! e^{-p\,t}\,\langle R^{2s}(t)\rangle_n \,dt\,
\rightarrow\,\frac {(2s)!}{(s-1)!}\cdot \frac 1p  \left(\frac
{v_0^2}{p^2}\right)^s\,\times \,
\,\,\,\,\,\,\label{momsn}\\
%
\begin{array}{lc}
\times\,  \left[\frac {p}{\gamma}\right]^{\,b+1}\int_0^1\frac
{(1-x)^{b-1}}{x^{\,b+1}}\,
   \left[\ln{\frac {1}{1-x}}
\right]^{s-1} dx\, & \,\,\,\texttt{if}\,\,\, s>b+1 \\
\times\,  \,\, \,\,\,\,\, \left(\frac {p}{\gamma}\right)^s
  \ln{\frac {\gamma}{p}} &
  \,\,\,\texttt{if}\,\,\, s=b+1  \\
\times\,\,\, \,\,\,\, \,\, \,\,\,\,\,\,  \left(\frac
{p}{\gamma}\right)^s
  \frac {\Gamma(s)\Gamma(b+1-s)}{\Gamma(b+1)} &
  \,\,\,\texttt{if}\,\,\, s<b+1  \\
\end{array} 
\nonumber
\end{eqnarray}
with\, $\,b=1/\alpha+n\,$\,. The first row here shows that high
enough moments are determined by not only the characteristic
diffusion lengths\, $\,\sqrt{2D_bt}\,$ and $\,\sqrt{2D_at}\,$ but
also by free-flight lengths\, $\,V_0t\,$ and $\,v_0\,$.

In the time domain, the asympt0tic behavior of fourth0order cumulamt
of BP's path what corresponds to (\ref{m4s})-(\ref{m4m}) is
\begin{eqnarray}
\langle R^{4}(t)\rangle_0 -3\langle R^{2}(t)\rangle_0^2\,
\rightarrow\, \,\,\,\,\,\,\,\,\,\,\,\,\,\,\,\,\,\, \,
\,\,\,\,\, \label{m4t}\\
\rightarrow\, 3\,(2D_b^2t)^2\, (\gamma t)^{1-M/m}\,
 \frac {2C(M/m)}{\Gamma(4-M/m)}\,
\,\,\, \,\,\,\, \texttt{if} \,\,\,\, \frac Mm
<1\,\,\,\,\nonumber\\
\rightarrow\, 3\,(2D_b^2t)^2\, \ln{(\gamma t)} \,\,\,\, \,\,\,
\,\,\,\, \texttt{if} \,\,\,\, \frac Mm
=1\,\,\,\,\nonumber\\
\rightarrow\, 3\,(2D_b^2t)^2\, \frac {m}{M-m} \,\,\,\, \,\,\,
\,\,\,\, \texttt{if} \,\,\,\, \frac Mm >1\,\,\,\,\nonumber
\end{eqnarray}

Interpreting these asymptotics as manifestation of low-frequency
fluctuations of BP's diffusivity, we see that in any case (i.e. at
any mass ratio $\,M/m\,$) that are non-ergodic fluctuations
represented by formally non-stationary random processes, in the sense
explained in \cite{i1}. Corresponding effective spectral densities of
the diffusivity fluctuations at frequencies $\,\ll \gamma\,$ are
\begin{eqnarray}
S_D(\omega) \rightarrow\, \frac {2\pi D_b^2}{\omega} \left[\frac
{\gamma}{\omega}\right]^{1-\frac Mm} \,\times\,\,\,\, \,\,\,\,\,
\label{m4f}\\
\times \, \frac {M/m}{(3-\frac Mm)(2-\frac
Mm)\cos{\left[ \frac {\pi}2\left(1-\frac Mm\right)\right]}} \,\,\,\,
\texttt{if} \,\,\,\, \frac Mm
<1\,\nonumber\\
S_D(\omega) \rightarrow\, \frac {\pi D_b^2}{\omega} \,\,\, \,\,\,
\,\,\,\, \texttt{if} \,\,\,\, \frac Mm =1\,
\, \nonumber\\
S_D(\omega) \rightarrow\, 2\pi D_b^2\,\delta(\omega)\,\frac m{M-m}
\,\,\,\, \,\,\, \,\,\,\, \texttt{if} \,\,\,\, \frac Mm
>1\,\nonumber
\end{eqnarray}
The latter expression means ``static'' fluctuations instead of  a
random process. pe, this is consequence of approximate character of
our consideration, while the exact one would change the factor
$\,\delta(\omega)\,$ to something like $\,\omega^{-1}
\ln^{\beta}{(\gamma/\,\omega)}\,$.

At last, consider asymptotic of sufficiently high-order moments.
Namely, at $\,s>b=1/\alpha +n+1=M/m+n+1\,$ from (\ref{moms}) we have
\begin{eqnarray*}
\int_0^\infty \!\! e^{-p\,t}\,\langle R^{2s}(t)\rangle_n \,dt\,
\rightarrow\,\frac {(2s)!}{(s-1)!}\cdot \frac 1p \left(\frac
{D_a}p\right)^b
\left(\frac {v_0^2}{p^2}\right)^{s-b}\,\times \,
\,\,\,\,\,\,\label{moms5}\\
\times\, \int_0^1\frac {(1-x)^{b-2}}{x^{b}} \left[\ln{\frac {1}{1-x}}
\right]^{s-1} dx\,\nonumber
\end{eqnarray*}
and correspondingly
\begin{eqnarray}
\langle R^{2s}(t)\rangle_n\, \rightarrow\, (D_at)^b\,
(v_0^2 t^2)^{s-b}\,\times \,
\,\,\,\,\,\, \,\,\,\,\,\, \,\,\,\,\,\,\label{hm}\\
\times\, \frac {(2s)!}{(s-1)!\, \Gamma(2s-b)} \int_0^1\frac
{(1-x)^{b-2}}{x^{b}} \left[\ln{\frac {1}{1-x}} \right]^{s-1}
dx\,\nonumber
\end{eqnarray}
The coefficient here can be estimated as
\begin{eqnarray}
\frac {\langle R^{2s}(t)\rangle_n}{(D_at)^b\, (v_0^2 t^2)^{s-b}}\,
<\,
\,\,\,\,\,\, \,\,\,\,\,\, \,\,\,\,\,\,\label{est}\\
<\, \frac {(2s)!}{(s-1)!\, \Gamma(2s-b)} \int_0^\infty\!
y^{s-b-1}\,(1+y)^b\,e^{-(b-1)y}\, dy\,\nonumber
\end{eqnarray}
Hence, for fixed $\,b=M/m+n+1\,$ and large enough $\,t\,$ we can
write
\[
\lim_{s\rightarrow\infty}\,\, \langle
R^{2s}(t)\rangle_n^{1/s}\,\lesssim\, \frac {v_0^2
t^2}{M/m+n}\,\,\,\,\,\,\,\,
\]
This means that the distributions $\,W_n(tl\bm{R})\,$ are sharply cut
off at characteristic ballistic-flight lengths  $\,|\bm{R}|\sim
t\sqrt{T/(M+nm)}\,$. In particular, the BP's path distribution
$\,W_0(tl\bm{R})\,$ is cut off at $\,|\bm{R}|\sim V_0 t\,$.

\section{Virial relations}

In conclusion let us concentrate on $\,W_n$'s dependence on the gas
density $\,\nu\,$. Obviously, it is the same as dependence on
$\gamma\,$, since $\,\gamma\propto\nu\,$ is the only
$\,\nu$-dependent parameter of our model, and therefore
$\,\gamma\partial/\partial\gamma = \nu\partial/\partial\nu\,$.

From Eq.\ref{erw1} it follows that
\begin{eqnarray}
\frac {W_{n+1}}{1+\alpha(n+1)} =\left[1+\frac
{1+p/\gamma}{1/\alpha+n+1}\,\, \nu\frac {\partial}{\partial \nu}\,
\right]\frac {W_{n}}{1+\alpha n}\,\,\,\,\label{vr}
\end{eqnarray}
In the long-range limit it turns into
\begin{eqnarray}
\frac {\overline{W}_{n+1}}{1+\alpha(n+1)} =\left[1+\frac
{1}{1/\alpha+n+1} \,\,\nu\frac {\partial}{\partial \nu}\,\right]
\frac {\overline{W}_{n}}{1+\alpha n}\,\,\,\,\label{vr1}
\end{eqnarray}
At that, in application to (\ref{us}), we cam make change $\,
\nu\partial/\partial\nu =- D_b\partial/\partial D_b-D_a
\partial/\partial D_a\,$.

In essence, formulas (\ref{vr})-(\ref{vr1}) represent straight
analogy of the ``virial relations'' investigated in
\cite{tmf,artmf,p3,jst,hs} on rigorous and most general level.
Special treatment of relations like (\ref{vr1}) will be done
elsewhere.

\begin{verbatim}
\end{verbatim}

\section{Resume}

In this paper, following \cite{i1} and \cite{p1}, we continued
approximate analysis of general equations of  the collisional
approximation to kinetics of spatially non-uniform gas \cite{i1} (see
also Sec.III above and \cite{i2}, \cite{p1} and \cite{hs1}). We
extended method of \cite{p1} (```diffusive approximation'') to the
case when the test ``Brownian'' particle (BP) has a mass different
from mass of the gas atoms, although posse the same friction, or
momentum relaxation rate, as the atoms. By the example of this
specific but interesting situation we demonstrated that both the
spectrum of low-frequency 1/f-type fluctuations in BP's diffusivity
and effective (time-smoothed) probability distribution of the
diffusivity essentially depend on the mass ratio of BP and atoms.
This means that probability distribution of BP's path depends on this
ratio, at that always possessing essentially non-Gaussian diffusive
long-range asymptotic.

In the case we considered under the approximation we used the
diffusivity 1/f-type, or ``flicker'', noise has the exponent equal to
or greater than unit. It remains unclear whether molecular random
walk in a gas (or in a liquid) can have exponents less than unit or
at least ``a little less'' as in the phenomenological theory
suggested in \cite{bk12,bk3} for charge transport (see also
\cite{i2}). To answer this question and consider more general
situations (first of all the case of different relaxation rates) we
should leave the just exploited approximation (since it presumes
presence of quasi-equilibrium inside the ``collisional clusters) and
start again from the equations (\ref{cfn}) \cite{i1} or even from
formally exact equations of molecular Brownian motion (see Sec.II and
\cite{tmf,artmf}).


\end{document}